\documentstyle[aps,epsf,rotate]{revtex}
\begin{document}
\draft
\twocolumn[\hsize\textwidth\columnwidth\hsize\csname @twocolumnfalse\endcsname
\title{Thermostatting by deterministic scattering}
\author{R. Klages,\cite{em} K. Rateitschak, G. Nicolis}
\address{Center for Nonlinear Phenomena and Complex Systems, 
Universit\'{e} Libre de Bruxelles, Campus Plaine CP 231, Blvd du
Triomphe, B-1050 Brussels, Belgium}
\date{\today}
\maketitle
\begin{abstract} 
We present a mechanism for thermalizing a moving particle by
microscopic deterministic scattering. As an example, we consider the
periodic Lorentz gas. We modify the collision rules by including
energy transfer between particle and scatterer such that the scatterer
mimics a thermal reservoir with arbitrarily many degrees of
freedom. The complete system is deterministic, time-reversible, and
provides a microcanonical density in equilibrium. In the limit of the
disk representing infinitely many degrees of freedom and by applying
an electric field the system goes into a nonequilibrium steady state.
\end{abstract}
]

Thermostatting is a mechanism by which the internal energy of a
many-particle system, and thus its temperature, are kept constant
although there is a flux of energy through the system as, e.g.,
induced by external fields, or by imposing temperature or velocity
gradients \cite{AT87}. In order to study nonequilibrium transport in
fluids by computer simulations, Evans, Hoover, Nos\'e and others
developed methods of thermostatting by introducing a fictitious
frictional force into the microscopic equations of motion 
\cite{EvMo90,MoDe98}, modeling the interaction of particles
with a thermal reservoir. This force is chosen to be velocity
dependent in a way that the (total or kinetic) energy of the particles
remains constant and that, in an equilibrium situation, the system
approaches microcanonical or canonical distributions of the phase
space variables. In contrast to stochastic thermostats
\cite{AT87,LeSp78,ChLe95}, this class of dynamical systems is
completely deterministic and time-reversible, making them suitable
models to study the connection between microscopic reversibility and
macroscopic irreversibility. This led to interesting new relations
between statistical physics and dynamical systems theory
\cite{EvMo90,MoDe98,DoVL}, especially between transport coefficients
and Lyapunov exponents \cite{MH87,PoHo88,ECM,Vanc}, and between
irreversible entropy production and phase space contraction
\cite{ChLe95,PoHo88,HHP87,Ch1,VTB97}. On the other hand, employing
these friction coefficients for creating nonequilibrium steady states
is physically somewhat obscure, because the equations of motion are
fundamentally modified. It has been shown that the resulting dynamical
systems belong to some new class of generalized Hamiltonian systems
\cite{EvMo90,MoDe98,Choq98}, but the problem still remains whether the
relations mentioned above rely on having specifically this class of
systems, or whether they are of a general nature
\cite{ChLe95,NiDa98,Gasp,MaHo92}. Much recent work
\cite{MaHo92,Mare97,TGN} has been devoted to deal with this question,
and to relate this way to model nonequilibrium states to the one
initiated by Gaspard, where nonequilibrium is induced by appropriate
boundary conditions in spatially extended systems \cite{Gasp,GN}.

One of the simplest examples of a dynamical system in which
thermostatting by a velocity-dependent friction coefficient has been
applied is the field driven periodic Lorentz gas. The classical
periodic Lorentz gas, where a point particle scatters elastically at
hard disks arranged on a triangular lattice, has been the subject of
many investigations and serves as a standard model in the field of
chaos and transport \cite{Gasp,MaHo92,Mare97,TGN,GN}. In the field
driven case an electric field acts on the moving particle and a
thermostat keeps the energy of the particle constant at all times
\cite{MoDe98,MH87,Vanc,Ch1,BarEC,LNRM95,DeGP95,DeMo97}. For unit
mass and unit charge of the particle the equations of motion read
$\mbox{\boldmath $\dot{r}$}=\mbox{\boldmath $v$}\:,\:\mbox{\boldmath
$\dot{v}$}=\mbox{\boldmath $\varepsilon$}-\zeta(\mbox{\boldmath
$v$})\mbox{\boldmath $v$}\: ,$ plus the geometric constraints induced
by the disks. $\mbox{\boldmath$r$}$ and $\mbox{\boldmath$v$}$
represent position and velocity of the moving particle,
$\mbox{\boldmath$\varepsilon$}$ is the electric field, and
$\zeta(\mbox{\boldmath $v$})$ stands for the friction coefficient
which, by requiring energy conservation, is determined to be
$\zeta(\mbox{\boldmath $v$})=\mbox{\boldmath $\varepsilon\cdot
v$}/v^2$. This set-up provides an example of a so-called Gaussian
thermostat \cite{EvMo90,MoDe98,DoVL}.

\begin{figure}
\epsfxsize=7cm
\centerline{\epsfbox{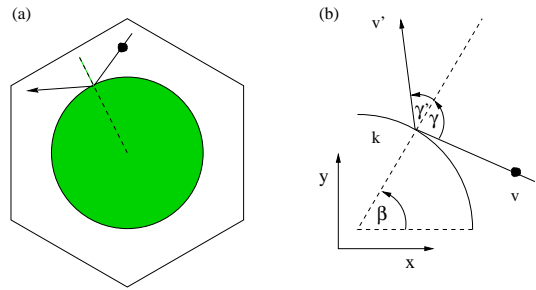}}
\vspace*{0.3cm} 
\caption{(a) Elementary cell of the periodic Lorentz gas on a
triangular lattice. (b) Definition of the relevant variables to
describe the collision process.}
\label{cell}
\end{figure}

In this Letter we propose an alternative mechanism of deterministic,
time-reversible thermostatting that does not appeal to such a
frictional force. We shall implement this mechanism by adapting the
geometry of the periodic Lorentz gas, in which it suffices to consider
one scatterer in an elementary cell supplemented by periodic boundary
conditions, as depicted in Fig.\ \ref{cell} (a). For the spacing
between two neighboring disks at disk radius $R=1$ we choose,
following the literature \cite{MH87,Vanc,DeGP95,DeMo97}, $w\simeq
0.2361$, ensuring that no particle can move collision-free for
infinite time. Fig.\ \ref{cell} (b) defines the relevant variables for
the collision process.  At the collision we express the velocity
$\mbox{\boldmath $v$}$ of the colliding particle in local polar
coordinates 

\noindent $(\gamma,v)$, where $\gamma$ is the angle of incidence and
$v$ the absolute value of the velocity. The dashed variables indicate
the respective values after the collision. We also introduce the angle
$\beta$, which determines the position of the colliding particle at
the disk. For an elastic collision, as in case of the original Lorentz
gas, one has $(\gamma,v)=(\gamma',v')$. In contrast to this, we
propose to include an energy transfer between particle and disk,
which, viewed in this way, serves as a momentum- and
energy-reservoir. We do this by introducing an additional velocity
variable $k$ associated to the disk and by allowing that
$\gamma\neq\gamma'$. We require that the energy is conserved at the
collision, $v^2+k^2=v'^2+k'^2=2E$, where $E$ is the total energy of
the system. Thus, the collision process in velocity space is still
effectively defined by the dynamics of two variables for which we take
$(\gamma,v)$. Using this setup, one can construct a model where the
disk rotates with $k$ as an angular velocity
\cite{RKN98}. By keeping the component of $v$ perpendicular to the
disk fixed and allowing only exchange of energy via the tangent
component, the collision process effectively reduces to the one of two
colliding masses on a line. Requiring energy and momentum conservation
yields for the scattering rules a two-dimensional piecewise linear
map. As a drastic simplification of such a model, but keeping
important dynamical properties like time-reversibility, a
deterministic dynamics, and the dynamical instability induced by the
disk geometry, we choose here our collision rules according to a
simple baker map \cite{DoVL,Ott,rotd}. We apply it on the respective
Birkhoff coordinate of the ingoing angle, $\sin|\gamma|$, as its
$x$-coordinate, and on $v$ in the range of $0\le v\le\sqrt{2E}$ as its
$y$-coordinate. To ensure that the system is time-reversible, the
forward baker acts if $0\le\gamma\le \pi/2$, and its inverse if
$-\pi/2\le\gamma<0$. The angle $\gamma'$ always goes to the respective
other side of the normal, as shown in Fig.\
\ref{cell} (b). For $\gamma\ge 0$ this gives
\begin{equation}
M(\sin|\gamma|,v)  = \left\{ 
\begin{array}{r@{,}l@{\:,\:}l}
(2\sin|\gamma| & v/2) & \sin|\gamma|\le 0.5 \\ 
(2\sin|\gamma| -1& (v+1)/2) & \sin|\gamma|> 0.5 \\ 
\end{array}
\right.
\label{eq:baker} 
\end{equation}
and {\em vice versa} for $\gamma<0$. As for $k'$, it is obtained from
energy conservation. To avoid any symmetry breaking induced by this
combination of forward and backward baker, we alternate their
application in $\gamma$ with respect to the position $\beta$ of the
colliding particle on the circumference\cite{alt}.

The above setting leads to a well-defined scattering system with three
degrees of freedom. As it stands, however, it does not satisfy the
microcanonical distribution, since the energy is not equipartitioned
between all degrees of freedom, corresponding to a distribution
$\rho(v_x,v_y,k)$ not being unifom on the energy shell
$v_x^2+v_y^2+k^2=2E$. We incorporate this essential feature by
amending the microscopic scattering rules, as given by the baker, in
the most straightforward way: As a starting point, we calculate the
projection of the microcanonical density
$\rho(v_x,v_y,k)=\delta(2E-v_x^2-v_y^2-k^2)$ onto $v$ yielding
$\rho(v)=v/\sqrt{2E(2E-v^2)}$. We want that our system approaches this
density in the long time limit. Let $\rho_{map}(v)$ be the probability
density for $v$ at the moment of the collision corresponding to a
respective Poincar\'e surface of section, in contrast to the
probability density of the time-continuous system, which we may denote
with $\rho_{cont}(v)$. During the free flight the particle cannot
change its velocity, and thus $\rho_{map}(v)$ and $\rho_{cont}(v)$ are
simply related via the average time the particle travels between two
collisions with velocity $v$. This average time plays the role of a
weighting factor leading to
\begin{equation}
\rho_{cont}(v)=\frac{c}{v} \rho_{map}(v) \quad , \label{eq:ttrafo} 
\end{equation}
where $c$ is a constant to be fixed by normalization. Thus, we want
that the map which determines the collision rules generates an
invariant velocity distribution $\rho_{cont}^*(v)\equiv\rho(v)$
corresponding to
\begin{equation} \rho_{map}^*(v)=\frac{2v^2}{\pi
E\sqrt{2E-v^2}} \quad .  
\end{equation} 
However, the invariant density of the baker map is simply
$\rho^*(x_B,y_B)=\rho^*(x_B)\rho^*(y_B)=1$. Therefore, we define a
conjugate map which produces the desired density by including a
transformation $y_B=Y(v)$, where $y_B$ is the actual baker
variable. This transformation must be continuous and invertible, as
defined by conservation of probability \cite{Ott},
\begin{equation} 
\rho^*(y_B)|dy_B|=\rho_{map}(v)|dv| \quad . \label{eq:pcons} 
\end{equation}
$Y(v)$ can then immediately be computed to
\begin{equation} 
Y(v)=-\frac{v}{\pi
E}\sqrt{2E-v^2}+\frac{2}{\pi}\arcsin\frac{v}{\sqrt{2E}}
\end{equation}
with $0\le v\le \sqrt{2E}\:,\:0\le Y(v)\le 1$. If we write
$x_B=X(\gamma)=\sin|\gamma|$, the full collision rules are thus given
by
\begin{equation}
(\gamma',v')=(X^{-1},Y^{-1})\circ M\circ(X(\gamma),Y(v))\quad .
\end{equation}
Computer simulations show that a Lorentz gas with these collision
rules is microcanonical in both its position and momentum coordinates
in phase space \cite{RKN98}.

We now inquire how the above ideas can be used to mimic the
interaction of a moving particle with a thermal reservoir. For this
purpose we associate arbitrarily many degrees of freedom to the disk
which could be related, e.g., to different lattice modes in a crystal,
as mechanisms for dissipating energy from a colliding particle. For
sake of simplicity we do not distinguish here between all the
individual velocities in the reservoir. Instead, we pretend that the
particle interacts instantaneously with {\em all} $(d-2)$ velocity
components $k_j$ of the reservoir via an absolute reservoir velocity
$k=\sqrt{k_1^2+k_2^2+\ldots+k_{d-2}^2}$, to which we identify the disk
velocity. We want that the projected densities of the accessible
variables $(v_x,v_y,k)$ be generated from the microcanonical
distribution of the full $d$-dimensional system. In particular, the
projection of the microcanonical distribution
$\rho_d(v_x,v_y,k_1,k_2,\ldots,k_{d-2})=
\delta(2E-v_x^2-v_y^2-\sum_{j=1}^{d-2}k_j^2)$
onto $\rho_d(v)$ can be calculated for $d>2$ to \cite{Kac59}
\begin{equation}
\rho_d(v)=(d-2) (2E)^{-\frac{d-2}{2}} v (2E-v^2)^{\frac{d-4}{2}} \quad
. \label{eq:ddens} 
\end{equation} 
Using the equipartition theorem $E/d=T/2$ with $T$ the temperature and
a Boltzmann constant $k_B=1$, and taking the limit $d\to\infty$, this
expression reduces to the Maxwellian distribution
$\rho_{\infty}(v)=v/T \exp(-v^2/(2T))$. Choosing
$\rho_{cont,d}^*(v)\equiv\rho_d(v)$ according to Eq.\ (\ref{eq:ddens})
and using Eq.\ (\ref{eq:ttrafo}) determines the corresponding density
$\rho_{map,d}^*(v)$ of the Poincar\'e section. The transformation
$Y_d(v)$ which yields $\rho_{map,d}^*(v)$ can then be calculated from
Eq.\ (\ref{eq:pcons}) \cite{RKN98}. In the limit of $d\to\infty$,
$Y_{\infty}(v)$ becomes especially simple and reads
\begin{equation}
Y_{\infty}(v)=-\sqrt{\frac{2}{\pi T}} v e^{-v^2/(2T)}+
\mbox{erf}\left(\frac{v}{\sqrt{2T}}\right)
\label{eq:dytrafo} 
\end{equation}
with $0\le v\le \infty\:,\:0\le Y(v)\le 1$. Computer simulations show
that a periodic Lorentz gas with these collision rules approaches
projected densities in $(v_x,v_y,k)$ which are identical to the ones
obtained from a uniform distribution on the energy shell in
$(v_x,v_y,k_1,k_2,\ldots,k_{d-2})$ \cite{RKN98}. The temperature $T$
of the equilibrium system is unambiguously defined via
equipartitioning of energy as it enters into Eq.\ (\ref{eq:dytrafo}).
Notice the complementarity between our approach and the procedure
adopted when using stochastic boundary conditions
\cite{LeSp78,ChLe95,WKN98}. 

We are now ready to set up a nonequilibrium situation by taking the
system as defined in equilibrium for $d\to\infty$ and by switching on
an electric field parallel to the $x$-axis. This field affects the
velocity of the moving particle. However, since the particle is a
small subsystem in a large reservoir, and since we have built in a
mechanism of equipartitioning of energy, the particle is still getting
thermalized by our scattering rules with a temperature determined by
the temperature $T$ of the reservoir. In fact, computer simulations
show that the system approaches a nonequilibrium steady state with
kinetic energy and conductivity fluctuating around constant mean
values
\cite{RKN98}. That such a nonequilibrium steady state exists according
to our scattering mechanism is the central result of our Letter.

\begin{figure}
\epsfysize=3.7cm
\centerline{\epsfbox{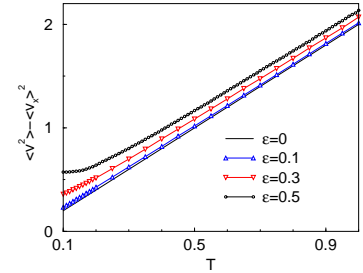}}
\caption{Relation between average velocity squared in
the frame moving with the current, $<v^2>-<v_x>^2$, and temperature
$T$ for the infinite dimensional model. Equipartitioning of energy
would imply $<v^2>-<v_x>^2=2T$.}
\label{entmp}
\end{figure}

\begin{figure}
\epsfysize=3.7cm
\centerline{\epsfbox{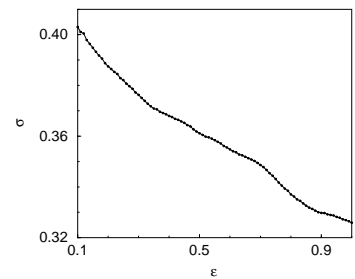}}
\caption{Conductivity $\sigma(\varepsilon)$ as it varies with field
strength $\varepsilon$. The curve consists of 90 data points, the
numerical uncertainty of each point is less than $5\cdot 10^{-4}$.}
\label{cond}
\end{figure}

In the following, we illustrate some important characteristics of this
state by results obtained from computer simulations. Fig.\ \ref{entmp}
demonstrates that for small enough field strength and high enough
temperature the energy of the system is still approximately
equipartitioned between $<v^2>-<v_x>^2$ and the reservoir. Fig.\
\ref{cond} depicts the conductivity $\sigma(\varepsilon)=<v_x>/\varepsilon$
with respect to the field strength $\varepsilon$. The strong decrease
of $\sigma(\varepsilon)$ indicates that the system is in a highly
nonlinear regime. Ohm's law may be suspected to hold only at very
small field strengths

\begin{figure}
\epsfysize=6.35cm
\centerline{\epsfbox{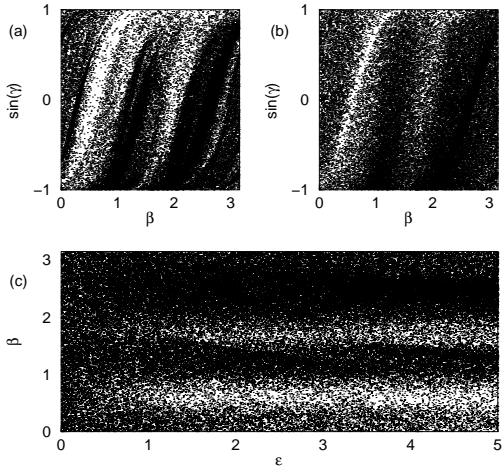}}
\caption{(a),(b) Poincar\'e section of $(\beta,\sin\gamma)$ defined in Fig.\
\ref{cell} at the moment of the collision for field strength
$\varepsilon=1$. In (a) the baker map of Eq.\ (\ref{eq:baker}) has been
used, in (b) the baker has been replaced by a random number
generator. (c) Poincar\'e section of $\beta$ at the moment of the
collision for varying field strength $\varepsilon$.}
\label{bifur}
\end{figure}

\noindent 
$\varepsilon \ll 0.1$ \cite{Ch1,vK71}, however, in this regime
reliable numerical results are difficult to get. The broadest
fluctuations on smaller scales are beyond numerical uncertainties and
may be reminiscent to the strong irregularities as they occur in the
conductivity of the Gaussian Lorentz gas
\cite{MH87,BarEC,LNRM95,DeGP95,DeMo97}. Unfortunately, it is not clear
how to compare the conductivities of these two models quantitatively,
since the choice of temperature in the Gaussian model is somewhat
ambiguous by a factor of two \cite{Ch1}. Fig.\
\ref{bifur} (a),(b) show Poincar\'e plots of $(\beta,\sin\gamma)$ at
the collisions. In (b) the deterministic baker has been replaced by a
random number generator such that the system completely loses its
memory at each collision. Fig.\ (a) indicates the existence of a
fractal attractor, analogously to the one found in the Gaussian
Lorentz gas \cite{MH87,DeGP95,HoMo89}, whereas in (b) the fractal
structure is lost due to the stochasticity of the boundary conditions.
Fig.\ \ref{bifur} (c) should be compared to the analogous diagram
obtained from the Lorentz gas with a Gaussian thermostat
\cite{MoDe98,LNRM95,LRM94}. In case of our model, there is no
indication of a pruning-induced ``bifurcation scenario'' or an ergodic
breakdown like in the Gaussian version. The same holds for other
choices of Poincar\'e sections in phase space.

It would be interesting to compare our approach in more detail to the
one based on frictional forces. For this purpose, a Lorentz gas with a
so-called Nos\'e-Hoover thermostat, where the particle velocity
fluctuates around a mean value \cite{MH87}, would be a more adequate
model to compare with, but we are not aware that such a model 
has been studied.  It would also be important to compute
Lyapunov exponents and fractal dimensions of the phase 

\noindent space densities
for our model. This would enable to test the status of recently
proposed connections between statistical physics and dynamical systems
quantities as discussed in Refs.\
\cite{Vanc,HHP87,Ch1,BarEC,LNRM95}. 

Our method has recently been applied to thermostat an interacting
many-particle system of hard disks under shear \cite{ChLe95}, and with
a temperature gradient, at the boundaries, where it leads to a
connection between deterministic thermostats and thermalization at
stochastic boundaries \cite{WKN98}. It would also be interesting to
use it as a simple model for inelastic scattering in granular
material, instead of employing velocity-dependent restitution
coefficients
\cite{AGZ98}.

We are indebted to P.Gaspard and Chr.Wagner for many important
hints. R.K.\ thanks the DFG for financial support, K.R.\ thanks the
European Commisssion for a TMR grant under contract no.\
ERBFMBICT96-1193.


\begin{references}
\bibitem{rotd}We note that our thermostatying mechanism works
as well with any other two-dimensional map, provided it is
area-preserving and chaotic \cite{WKN98}.
\end{references}
\end{document}